\documentclass{phb-proc4-auth}
\usepackage{graphicx}
\usepackage{amsmath}
\usepackage{amssymb}
\usepackage{times}

\graphicspath{{.}{../sceshol/}}

\begin{document}

\begin{frontmatter}

  \title{Delocalisation transition in chains with correlated disorder}

  \journal{SCES '04}
  
  \author[hgw]{Gerald Schubert},
  \author[syd]{Alexander Wei{\ss}e},
  \author[hgw]{Holger Fehske\corauthref{1}}
  \address[hgw]{Institut f\"ur Physik,
    Ernst-Moritz-Arndt Universit\"at Greifswald, 17487 Greifswald,
    Germany}
  \address[syd]{School of Physics, The University of
    New South Wales, Sydney NSW 2052, Australia} 

  \corauth[1]{Corresponding Author:  Phone: +49-3834-86-4760, 
    Fax +49-3834-86-4701, Email: fehske@physik.uni-greifswald.de}

  \begin{abstract}
    We show that in the one-dimensional (1D) Anderson model long-range
    correlations within the sequence of on-site potentials may lead to
    a region of extended states in the vicinity of the band centre,
    i.e., to a correlation-induced insulator-metal transition.  Thus,
    although still disordered, the 1D system can behave as a
    conductor.
  \end{abstract}
  \begin{keyword}
    disordered electron systems\sep Anderson metal-insulator transition 
  \end{keyword}

\end{frontmatter}

In the context of electronic transport in quasi-1D binary solids,
polymer chains or biological molecules, recently systems with
correlations in the sequence of on-site energies attracted increasing
attention~\cite{CBIS02}.  The examination of models with correlated
disorder is also challenging from a theoretical point of view.  In 1D
noninteracting electron systems with independent random atomic
potentials all states are exponentially localised at any amount of
disorder~\cite{Be73}. Internal correlations within the potential
landscape, however, might cause a breakdown of the Anderson
localisation phenomenon in 1D and 2D~\cite{Fl89,dML98}.  Such
long-range correlated random sequences without any intrinsic scale are
observed in several stochastic processes in nature~\cite{PMB96}.  One
of their characteristics is the power law decay of the Fourier
transform of the two-point correlation function with the wavenumber of
the random fluctuations,
$\mathcal{F}(\langle\epsilon_i\epsilon_j\rangle)\sim 1/k^\alpha$.
Based on these findings de Moura and Lyra~\cite{dML98} proposed an
Anderson type model, $H = \sum_j \epsilon_j {c}_j^{\dag} {c}_j^{} - t
\sum_{\langle ij \rangle} [{c}_i^{\dag} {c}_j^{} + \text{H.c.}]$, with
the following ansatz for the on-site potentials,
\begin{equation}\label{Corr_Dis}
  \epsilon_j = \nu(\alpha) \sum_{k=1}^{N/2} k^{-\alpha/2}
  \cos(2\pi j k/N + \phi_k)\,,
\end{equation}
where the $N/2$ random phases $\phi_k$ are uniformly distributed in
the interval $[0,2\pi]$ and $\alpha$ controls the strength of the
correlation. The case $\alpha=0$ corresponds to strongest disorder and
is almost equivalent to the uncorrelated Anderson model with Gaussian
distributed on-site potentials. In order to obtain comparable results
each disordered sequence is normalised to variance $\sigma^2 = 1$,
i.e. $\nu(\alpha) = (\sum_{k=1}^{N/2} k^{-\alpha} / 2)^{-1/2}$. This  
ensures that both disorder and bandwidth remain finite in the
thermodynamic limit.

In the theoretical investigation of disordered systems it turned out
that distribution functions for the random quantities of interest take
the centre stage~\cite{AAT73}.  Of particular importance is the
probability density $p(\rho_i(E))$ of the local density of states
(LDOS)
\begin{equation} \label{LDOS}
  \rho_i(E) = \sum_{n=1}^{N} |\psi_{n,i}|^2 \,\delta(E-E_n)\,.
\end{equation}
For a given energy $E$, $\rho_i(E)$ measures the local amplitude of
the wave function at Wannier site $i$ and therefore contains direct
information about the localisation properties.  Contrary to the mean
(arithmetically averaged) density of states, $\rho_{\rm av}(E) =
\langle\rho_i(E)\rangle$, the LDOS becomes critical at the
localisation transition. Concurrently $p(\rho_i(E))$ was found to have
essentially different properties for localised and extended
phases~\cite{BAF04}.  Of course, the study of entire distributions is
a bit inconvenient, and for practical calculations the so-called
``typical DOS'' (geometric average) $\rho_{\rm ty}(E) = \exp\langle
\ln\rho_i(E)\rangle$ is frequently used. Since $\rho_{\rm ty}(E)$
vanishes at the Anderson transition it acts as a kind of ``order
parameter''.

\begin{figure}
  \centering 
  \includegraphics[width=\linewidth,clip]{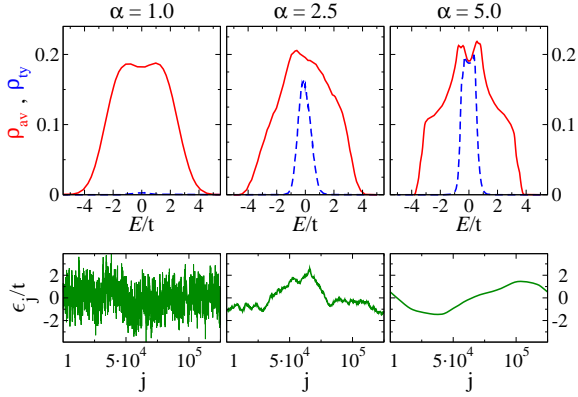}
   \caption{Average (solid) and typical (dashed) DOS for the
    Anderson model with correlated disorder on a linear chain with
    $125000$ sites and different correlation strengths $\alpha$.  The
    typical DOS was calculated at $K_s=32$ sites for $K_r=32$
    realizations of disorder using $32768$ Chebyshev moments, whereas
    for the average DOS only $256$ Chebyshev moments and a random
    initial vector were used. The lower panels show one characteristic
    sequence of atomic potentials $\epsilon_j$ for each value of
    $\alpha$.}\label{Corr_vgl}
\end{figure}

Numerically the LDOS distribution can be easily calculated with the
kernel polynomial method~\cite{SWF03}.  The corresponding arithmetic
and geometric averages $\rho_{\text{av}}$ and $\rho_{\text{ty}}$ are
shown in Fig.~\ref{Corr_vgl}.  In contrast to the uncorrelated
Anderson model we observe the following effects: (i) most notably,
while for small values of $\alpha$ the typical DOS vanishes for the
entire band, with stronger correlation a region of extended states
appears near the band centre; (ii) for larger values of $\alpha$ a dip
in $\rho_{\text{av}}$ develops at the band centre, which might be a
signature of the 1D DOS of the completely ordered system that is
expected for infinitely strong correlation; (iii) even though on
average the model is symmetric with respect to $E\rightarrow-E$ in
Fig.~\ref{Corr_vgl} the averaged density of states shows a noticeable
asymmetry for intermediate values of $\alpha$.  We fear that, due to
the longer ranged fluctuations within the on-site potentials for
increasing $\alpha$, the considered rather large ensemble
($32\times32$ realizations) is still insufficient to achieve a proper
statistics.

\begin{figure}
  \centering 
  \includegraphics[width=0.95\linewidth,clip]{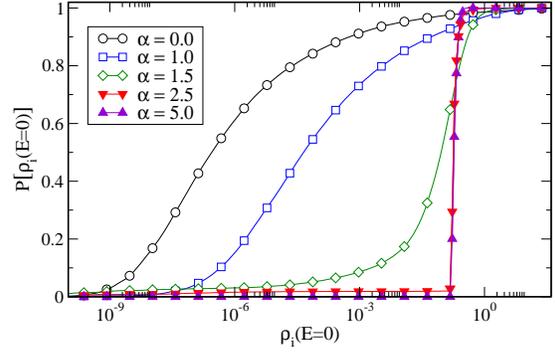}
  \caption{Probability distribution $P(\rho_i)$ in the band-centre
    for different correlation strengths $\alpha$ (again $N=125000$,
    $M=32768$, $K_r\times K_s = 32\times 32$).}
 \label{Corr_matlab}
\end{figure}

\begin{figure}
  \centering 
  \includegraphics[width=.95\linewidth,clip]{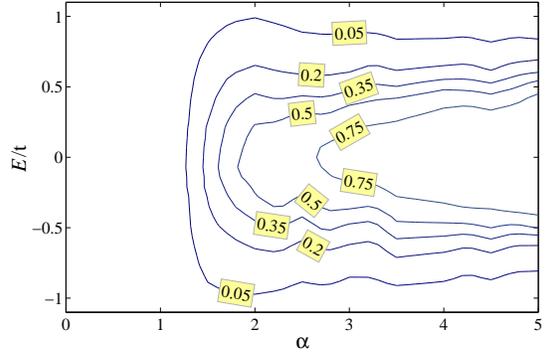}
  \caption{Contour plot of $R$ in the correlation-energy plane for the
    Anderson model with correlated disorder on a linear chain with
    $125000$ sites. In the calculation $32768$ Chebyshev moments 
    and $K_r\times K_s = 128\times 32$
    realizations were used.}\label{Prob_dis}
\end{figure}

Figure~\ref{Corr_matlab} illustrates the behaviour of the probability
distribution $P(\rho_i)=\int_0^{\rho_i} p(\rho_i^{\prime}) d
\rho_i^{\prime} $. At small values of $\alpha$, $P(\rho_i)$ reflects
the characteristics of the LDOS distribution of localised states, i.e.
large weight on small values of $\rho_i$ and a long tail.  The
jump-like increase of $P(\rho_i)$ for $\alpha \gtrsim 2$ signals the
occurrence of a very narrow symmetric distribution centred around
$\rho_{\text{av}}$ and describes extended states.  
Finally
Fig.~\ref{Prob_dis} shows contours of the ratio
$R(E)=\rho_{\text{ty}}(E)/\rho_{\text{av}}(E)$.  Apparently all states
become localised for $\alpha \lesssim 2$, whereas the width of the
region of extended states saturates for $\alpha \gtrsim 5$.

In conclusion, we analysed the effect of impurity correlations on the
localisation properties of 1D electron systems and showed that our
model exhibits an insulator metal transition with increasing
correlation strength.

Work was supported by NIC J\"ulich, HLRN Berlin-Hannover,
and DFG under SPP 1073 (FE 398/1-3). 



\end{document}